\documentclass{amsart}
\usepackage{amssymb, amsmath, epsfig}
\usepackage{amscd}

\newtheorem{thm}{Theorem}
\newtheorem{defn}{Definition}
\newtheorem{prop}[thm]{Proposition}
\newtheorem{lem}{Lemma}

\newtheorem{remark}{Remark}

\newcommand{\q}{\mathbf q}

\DeclareMathOperator{\diag}{diag}

\def\beq{\begin{equation}}                     %
\def\eeq{\end{equation}}                       %
\def\bea{\begin{eqnarray}}                     
\def\eea{\end{eqnarray}}                       
\def\tr{\hbox{\footnotesize Tr}}

\def\0 {\nonumber}

\newlength{\intwidth}
\DeclareRobustCommand{\fpint}[2]
   {\mathop{%
      \text{%
        \settowidth{\intwidth}{$\int$}%
        \makebox[0pt][l]{\makebox[\intwidth]{$-$}}%
        $\int_{#1}^{#2}$}}}

\begin{document}
\title[Chern-Simons theory on $L(p,q)$ and Gopakumar-Vafa duality]{Chern-Simons theory on $L(p,q)$ lens spaces and
Gopakumar-Vafa duality
}

\author[A. Brini]{Andrea Brini}
\address{Mathematical Physics sector, International School for Advanced Studies (SISSA/ISAS), Via Beirut 2-4, I-34014, Trieste, Italy and INFN, sezione di Trieste}
\email{brini@sissa.it, tanzini@sissa.it}
\author[L. Griguolo]{Luca Griguolo}
\address{Dipartimento di Fisica, Universit\`a di Parma and INFN Gruppo Collegato di Parma, Viale G.P. Usberti 7/A, 43100 Parma, Italy}
\email{griguolo@fis.unipr.it}
\author[D. Seminara]{Domenico Seminara}
\address{Dipartimento di Fisica, Universit\`a di Firenze and INFN Sezione di Firenze, Via G. Sansone 1, 50019 Sesto Fiorentino, Italy}
\email{seminara@fi.infn.it}
\author[A. Tanzini]{Alessandro Tanzini}
\thanks{SISSA Preprint 56/2008/FM}
\subjclass[2000]{81T45 (primary), 81T30, 57M27, 17B37, 14N35}
\keywords{Chern-Simons theory, Gopakumar-Vafa, large N duality, open-closed duality, topological strings, geometric transitions, random matrices.}
\begin{abstract}
We consider aspects of Chern-Simons theory on $L(p,q)$ lens spaces and its relation with matrix models and topological string theory on Calabi-Yau threefolds, searching for possible new large $N$ dualities via geometric transition for non-$SU(2)$ cyclic quotients of the conifold. To this aim we find, on one hand, a useful matrix integral representation of the $SU(N)$ $CS$ partition function in a generic flat background for the whole $L(p,q)$ family and provide a solution for its large $N$ dynamics; on the other, we perform in full detail the construction of a family of would-be dual closed string backgrounds via conifold geometric transition from $T^*L(p,q)$. We can then explicitly prove the claim in \cite{Auckly:2007zw} that Gopakumar-Vafa duality in a fixed vacuum fails in the case $q>1$, and briefly discuss how it could be restored in a non-perturbative setting.
\end{abstract}
\maketitle

\section{Overview}

After the seminal work of Witten \cite{Witten:1988hf}, Chern-Simons
($CS$) theory has been deeply studied both in Mathematics and Physics.
A most attractive property of this topological
field theory is its large $N$ duality with the $A$-model topological
string as
discovered by Gopakumar and Vafa in \cite{Gopakumar:1998ki}.
From the Physics viewpoint, this is a concrete realization
of 't Hooft's intuition \cite{'tHooft:1973jz} that the large $N$ Feynman expansion of a gauge theory
with $U(N)$ structure group can be recast
as a perturbative expansion of closed oriented strings in a suitable background; mathematically, such duality is a precise
(and amazing!) correspondence between two seemingly very
unrelated mathematical objects, namely knot invariants
and (relative) Gromov-Witten invariants.
This duality is realised through a particular kind of {\it geometric transition},
called {\it conifold transition}, which plays a relevant r\^ole in the study
of the moduli space of Calabi-Yau three-folds ($CY3$)
(see {\it e.g.} \cite{Grassi:2002tz, Auckly:2007zw,Rossi:2004eq} for reviews).

\noindent Let us recall the basic features of Gopakumar-Vafa ($GV$) duality. We have the following
\begin{prop}[Witten, \cite{Witten:1992fb}]
Let $M$ be a closed smooth 3-manifold such that $T^*M$ is a Calabi-Yau threefold. Then the open topological $A$-model on $T^*M$, with $N$ Lagrangian branes wrapping $M$, is equivalent to $U(N)$ Chern-Simons theory on $M$.
\label{wittencs}
\end{prop}
\begin{prop}[Gopakumar-Vafa, \cite{Gopakumar:1998ki}]
The topological open $A$-model on $T^*S^3$ with $N$ $A$-branes wrapping the base $S^3$ is equivalent at large $N$ to the closed topological $A$-model on $\mathcal{O}_{\mathbb{P}^1}(-1)\oplus\mathcal{O}_{\mathbb{P}^1}(-1)$.
\label{GV}
\end{prop}\noindent
The closed string target space $\mathcal{O}_{\mathbb{P}^1}(-1)\oplus\mathcal{O}_{\mathbb{P}^1}(-1)$ and the open string one $T^*S^3$ are related by topological surgery - a birational contraction plus a complex deformation of a nodal singularity. This is what goes under the name of ``{\it (conifold) geometric transition}''. As it stands, the content of Proposition \ref{GV} is striking and somewhat mysterious: a topological invariant of a 3-manifold - the $CS$ partition function, or Reshetikin-Turaev-Witten invariant of $S^3$ - is also a generating function of symplectic invariants of a non-compact K\"ahler manifold - its Gromov-Witten potential. It would be then extremely interesting to see if one could find new examples of such a duality along the same lines, replacing $S^3$ with a generic 3-manifold and engineering the geometric transition in such a case. Indeed, one expects on general grounds the duality to hold for Chern-Simons theories on rational homology spheres, but actually very few examples beyond $S^3$ are known:\\

\noindent {\bf Open problem.} {\it
Find when the Reshetikin-Turaev-Witten invariant of a compact smooth 3-manifold $M$ is equal, in suitable coordinates, to the all genus Gromov-Witten potential of an algebraic threefold $X_M$ obtained by geometric transition from $T^*M$: that is,  $X_M$ is given by a complex deformation of $T^*M$ to a normal variety, followed by a birational resolution.} \\

\noindent This issue has been addressed in the case of $\mathbb{Z}_p\subset SU(2)$ cyclic quotients of $S^3$:

\begin{prop}[\cite{Halmagyi:2003mm}]
Let $M=L(p,1)$. The Hori-Vafa mirror curve and differential of the family of $CY3$ obtained by geometric transition from $T^*M$ coincide with the spectral curve and resolvent of the $L(p,1)=M$ Chern-Simons matrix model \cite{Marino:2002fk}.
\label{HYO}
\end{prop}\noindent

This has been achieved through a detailed study of the random matrix representation for the Chern-Simons path integral, originally obtained by Mari\~no
in \cite{Marino:2002fk}. An early confirmation of the above assertion was provided in \cite{Aganagic:2002wv} for $p=2$, by matching the 't Hooft expansion of the Chern-Simons 2-matrix model with the solutions of the Picard-Fuchs system at the orbifold point of $K_{\mathbb{P}^1 \times \mathbb{P}^1}$. Notice that, assuming\footnote{There is presently a strikingly large body of evidence of the validity of mirror symmetry for toric $CY3$; see also \cite{mirror}
for rigorous mathematical proofs.}
the validity of local mirror symmetry for toric $CY3$, Proposition \ref{HYO} implies $GV$ duality for $L(p,1)$ lens spaces at least in genus zero; actually, recent progress  \cite{Bouchard:2007ys} strongly suggests that the spectral data contain the full structure of the $B$-model on toric $CY3$ at all genera, in which case the work \cite{Halmagyi:2003mm} would become automatically an all-genus proof. \\

However, it has been suspected that one could hardly go further along this direction for generic $M$. Actually the most exhaustive and detailed monograph on the subject contains the following \\

\noindent {\bf Claim 1 (\cite[\S 7.3]{Auckly:2007zw}).} {\it Proposition \ref{HYO} is false when $M=L(p,1)$ is replaced by the generic lens space $M=L(p,q)$ for $q>1$.}
\\

In this note we will perform a detailed analysis of the whole family $M=L(p,q)$, thus including cyclic subgroups not contained in $SU(2)$, along the lines of \cite{Halmagyi:2003mm} and assuming mirror symmetry. The case  $q>1$, in which the lens space is not a $U(1)$ bundle on $S^2$, appears to be much harder as most of the features of the $q=1$ case, like the mirror realization of \cite{Aganagic:2002wv} 
become either unclear or simply are not there. 
To this aim, in section \ref{closed} we will work out explicitly the conifold transition for the case at hand
 and  obtain a class of would-be large $N$ duals $\widehat{\mathcal{X}_{p,q}}$ as well as their Hori-Vafa mirror curves, correcting {\it en passant} a few claims in the literature about the impossibility of performing such a geometric transition in the $L(p,q)$ case preserving at the same time the $CY$ condition; in section \ref{open} we will present a matrix integral representation for the Chern-Simons $U(N)$ partition function on $L(p,q)$ in a fixed flat connection background and consider its large $N$ expansion as governed by a spectral curve and a resolvent.
We will find disagreement for $q>1$ with the $B$-model spectral data obtained after geometric transition from $T^*L(p,q)$; under the assumption of validity of local mirror symmetry for toric $CY3$, this proves the claim above.\\

Even though our proof of Claim 1 might appear to be an obstruction to the program of extending the duality of \cite{Gopakumar:1998ki} to more general backgrounds, let us outline
two possible avenues of further investigation which we think might lead to the solution of the puzzle for the case under scrutiny. \\

One possible way out is to regard $GV$ duality as an identity between the full $CS$ partition on a 3-manifold $M$ and some suitable {\it non-perturbative} definition of the $A-$model on the $CY3$ obtained through conifold geometric transition from $T^*M$. Indeed, as first advocated in\footnote{See also \cite{Eynard:2008yb, Eynard:2008he} for related work on background independence and \cite[\S 6.3]{Marino:2008ya}, \cite[\S 5.2]{Eynard:2008he} for a discussion precisely about the case of topological strings with a $L(p,1)$ Chern-Simons matrix model representation.} \cite{Marino:2007te}, a proper non-perturbative definition of the $A-$model on toric target spaces with a dual matrix integral description should be given in terms of a filling fraction independent sum over multi-instanton sectors. This would be dual to the proper definition of the Reshetikin-Turaev-Witten invariant as a sum over flat connections. \\ A second possibility, hinted at by the geometric picture arising in the discussion of section 2, might consist in a refinement of the notion of ``orbifold of the $GV$ duality for $S^3$'' in order to properly encompass the case of the generic lens space. 
Indeed, for $1<q<p-1$ the cyclic group does no longer act fiberwise on the resolved conifold, giving rise to
an orbibundle over a rational curve with marked points (see Remark \ref{orbicurve}). This new feature with respect to the $q=1$ case definitely begs for further understanding, in order to clarify the correct formulation of $GV$ duality in this case as well as its possible relation with a (suitably twisted) Gromov-Witten theory of orbicurves.
%
%
\\

We hope to address both this issues in future work.\\

\section{The closed string side: conifold transition for $T^*L(p,q)$}
\label{closed}
\subsection{Geometric transition}
According to Proposition \ref{wittencs} and \ref{GV}, the $GV$ duality for the case of the generic $L(p,q)$ lens space should be realized in two steps:
\begin{enumerate}
\item a complex deformation of $\widehat{\mathcal{X}_{p,q}}\equiv T^*L(p,q)$ to a normal variety $\mathcal{X}_{p,q}$ (a suitable $\mathbb{Z}_p$ quotient of the singular conifold);
\item a complete crepant resolution $\overline{\mathcal{X}_{p,q}}$ of the latter.
\end{enumerate}The first step is realized as follows: let us recall that, from \cite[Theorem 1.6]{Grassi:2002tz}, the cotangent bundle to the 3-sphere $T^*S^3$ is diffeomorphic to a smooth hypersurface in $\mathbb{A}^4$
\beq
\label{conifoldeq}
xy -zt =\mu
\eeq
which is a complex structure deformation of a conifold singularity. The base $S^3$ is the real locus $y=\bar x$, $t=-\bar z$
\beq
|x|^2 + |z|^2 = \mu
\label{S3inT*S3}
\eeq
Now, consider the $\mathbb{Z}_p$ action
\beq
\begin{array}{cccccc}
\mathbb{Z}_p & \times & \mathbb{C}^4 & \to & \mathbb{C}^4 \\
\omega & & (x,y,z,t) & \to & (\omega x,\omega^{-1} y, \omega^{q} z, \omega^{-q} t)
\label{zpaction}
\end{array}
\eeq
where $\omega = e^{2\pi i/p}$, $1\leq q <p$, $(p,q)=1$; the orbit manifold restricted to (\ref{S3inT*S3}) is a $L(p,q)$ lens space. At first sight, using the same coordinatization as \cite{Grassi:2002tz}, the cyclic group acts both on the fibers and on the base of $T^*S^3$, thus yielding something a priori different from an $\mathbb{R}^3$-bundle over $L(p,q)$. However we have the following simple
\begin{lem}
The orbit space of (\ref{zpaction}) restricted to (\ref{conifoldeq}) is smoothly diffeomorphic to $T^*L(p,q)$.
\end{lem}
{\bf Proof.} Introduce the new set of variables $w_i=q_i+ip_i$
\beq
\begin{array}{ccccccc}
w_1 &=& (z_1+z_3)/2 &\qquad & w_2 &=& i(-z_1+z_3)/2 \\
w_3 &=& (z_2+z_4)/2 &\qquad & w_4 &=& i(-z_2+z_4)/2
\end{array}
\eeq
In this coordinates, (\ref{conifoldeq}) is the locus in $\mathbb{R}^8$ described by
\beq
\sum_{j=1}^4 q_j^2 - p_j^2=\mu, \qquad \sum_{j=1}^4 q_j p_j=0
\label{cots3}
\eeq
Consider then the change of variables
\beq
\begin{array}{ccccccc}
\tilde p_1 &=& q_1 p_1 + q_2 p_2, & \qquad & \tilde p_2 &=& q_1 p_2 - q_2 p_1 \\
\tilde p_3 &=& q_3 p_3 + q_4 p_4, & \qquad & \tilde p_4 &=& q_3 p_4 - q_4 p_3
\end{array}
\eeq
and $\tilde q_i = q_i$, $i=1, \dots, 4$. The change of variables for $\mu >0$ is nonsingular everywhere in the set defined by (\ref{cots3}), which is then rewritten as
\beq
\label{invts3}
\sum_{i=1}^4 \tilde q_i^2 = \mu, \qquad \tilde p_1 + \tilde p_3 = 0
\eeq
The $\mathbb{Z}_p$ action is now represented on the tilded $\mathbb{R}^8$ in the form:
\beq
\left(
\begin{array}{c}
\tilde q_1 \\
\tilde q_2 \\
\tilde q_3 \\
\tilde q_4
\end{array}
\right) \to
\left(
\begin{array}{cccc}
\cos{2\pi/p} & \sin{2\pi/p} & 0 & 0 \\
-\sin{2\pi/p} & \cos{2\pi/p} & 0 & 0 \\
0 & 0 & \cos{2\pi q/p} & \sin{2\pi q/p} \\
0 & 0 & -\sin{2\pi q/p} & \cos{2\pi q/p}
\end{array}
\right)
\left(
\begin{array}{c}
\tilde q_1 \\
\tilde q_2 \\
\tilde q_3 \\
\tilde q_4
\end{array}
\right)
\eeq
\beq
\left(
\begin{array}{c}
\tilde p_1 \\
\tilde p_2 \\
\tilde p_3 \\
\tilde p_4
\end{array}
\right) \to
\left(
\begin{array}{c}
\tilde p_1 \\
\tilde p_2 \\
\tilde p_3 \\
\tilde p_4
\end{array}
\right)
\eeq
realizing therefore the $\mathbb{Z}_p$ quotient (\ref{zpaction})  of the deformed conifold as a trivial $\mathbb{R}^3$-bundle over $L(p,q)$. By Stiefel's theorem \cite{Stiefel}, the latter being an orientable three-manifold, there exists a (strong) $C^\infty$ bundle isomorphism mapping $\mathbb{R}^3 \times L(p,q)$ to $T^*L(p,q)=:\widehat{\mathcal{X}_{p,q}}$.
\begin{flushright}$\square$\end{flushright}
With the algebraic realization (\ref{conifoldeq}), (\ref{zpaction}) of $\widehat{\mathcal{X}_{p,q}}$ at hand it is straightforward to perform the second step of the transition. As in the $S^3$ case, the $\mu$ parameter measures the size of the lens space and sending $\mu$ to zero amounts to deforming $\widehat{\mathcal{X}_{p,q}}$ to the singular variety $\mathcal{X}_{p,q}$, where the Lagrangian null section $L(p,q)$ has shrunk to zero size. We have the following
\begin{thm}
The singular variety $\mathcal{X}_{p,q}$, obtained as the orbit space of (\ref{zpaction}) inside (\ref{conifoldeq}) with $\mu=0$, is a toric variety with trivial canonical sheaf, $K_{\mathcal{X}_{p,q}}\simeq\mathcal{O}_{\mathcal{X}_{p,q}}$.
\end{thm}
{\bf Proof.} For $q=1$ the theorem was proven in \cite{Halmagyi:2003mm}, where the authors exploited the fact that $\mathcal{X}_{p,1}$ is obtained from the resolved conifold (a rank 2 bundle over $S^2$ as in proposition \ref{GV}) by quotienting a fiberwise-acting $\mathbb{Z}_p$ group and ``blowing-down'' the base $S^2$. For $q>1$, though, the $\mathbb{Z}_p$ group does no longer act fiberwise and we have to deal with it in a different way.

By definition\footnote{See \cite{Fulton}  for an introduction to toric geometry.}, we have to prove that $\mathcal{X}_{p,q}$ contains an algebraic three-torus as an open subset effectively acting through an extension of its obvious action on itself. This is identified as follows: the singular conifold $\mathcal{X}$, as an affine variety
$$\mathcal{X}:=\hbox{Spec} \frac{\mathbb{C}[x,y,z,t]}{\{xy-zt\}}$$
is toric with torus action given by
\bea
\label{3toruscf}
(\mathbb{C}^*)^3 & \stackrel{j}{\hookrightarrow} & \mathcal{X} \nonumber \\
(t_1, t_2, t_3) & \to & (t_1, t_2, t_3, t_1 t_2 t_3^{-1})
\eea
This action descends to an action on the orbifolded conifold $\mathcal{X}_{p,q}$ by (\ref{zpaction})
\bea
(\mathbb{C}^*)^3/\mathbb{Z}_p & \stackrel{\tilde j}{\hookrightarrow} & \mathcal{X}_{p,q}
\eea
Proving that $\mathcal{X}_{p,q}$ is toric therefore amounts to find explicitly an isomorphism $\pi : (\mathbb{C}^*)^3/\mathbb{Z}_p \mapsto (\mathbb{C}^*)^3$
\beq
\label{exseq}
0  \rightarrow \mathbb{Z}_p \stackrel{i}{\rightarrow} (\mathbb{C}^*)^3 \stackrel{\pi}{\rightarrow} (\mathbb{C}^*)^3 \rightarrow 0
\eeq
where the injection $i$ is dictated by (\ref{zpaction}) to be
\beq
\begin{array}{cccc}
i : & \mathbb{Z}_p &  \hookrightarrow &  (\mathbb{C}^*)^3 \\
& \omega & \mapsto & (\omega, \omega^{-1}, \omega^q)
\end{array}
\eeq
and by (\ref{exseq}) we can write for $\pi$
\beq
\begin{array}{cccc}
\pi : & (\mathbb{C}^*)^3 &  \hookrightarrow &  (\mathbb{C}^*)^3 \\
& (t_1, t_2, t_3) & \mapsto & (t_1^p, t_1 t_2, t_1^q t_3^{-1})
\end{array}
\eeq
The three-torus inside the quotient of the conifold by the action (\ref{zpaction}) is then identified by
\bea
\label{3toruspq}
(\mathbb{C}^*)^3 & \stackrel{\tilde j \circ \pi^{-1}}{\hookrightarrow} & \mathcal{X}_{p,q} \nonumber \\
(t_1, t_2, t_3) & \to & (t_1^{1/p}, t_1^{-1/p}t_2, t_1^{q/p}t_3^{-1}, t_1^{-q/p}t_2 t_3)
\eea
From (\ref{3toruspq}) we can read off the dual cone as the real tetrahedron spanned by
\beq
a_1 = \left(\begin{array}{c} 1/p \\ 0\\0\end{array}\right)  \quad  a_2 = \left(\begin{array}{c} -1/p \\ 1\\0\end{array}\right) \quad a_3 = \left(\begin{array}{c} q/p \\ 0\\-1\end{array}\right) \quad a_4 = \left(\begin{array}{c} -q/p \\ 1\\1\end{array}\right)
\eeq
The fan is then obtained by taking the inward pointing normal to each facet, normalized in such a way to hit the first point on the $\mathbb{Z}^3$ lattice. Modulo an automorphism of the lattice, we thus get that the rays of the fan of $\mathcal{X}_{p,q}$ are given by
\beq
\label{fanpq}
b_1 = \left(\begin{array}{c} 0 \\ 1\\q\end{array}\right)  \quad  b_2 = \left(\begin{array}{c} 0 \\ 1\\q+1\end{array}\right) \quad b_3 = \left(\begin{array}{c} p \\ 1\\1\end{array}\right) \quad b_4 = \left(\begin{array}{c} p \\ 1\\0\end{array}\right)
\eeq
and the fan consists of a single cone generated by $b_i$, $i=1 \dots 4$. Notice that the tip of the rays all lie in an an affine hyperplane, namely $y=1$, thus implying triviality of the canonical class \cite{Fulton}. The theorem is proved.
\begin{flushright}$\square$\end{flushright}

\begin{remark}\label{orbicurve}
It is instructive to point out an interesting new geometrical fact in the $1<q<p-1$ case. Let us consider the orbifold of the resolved conifold
geometry $\mathcal{O}_{\mathbb{P}^1}(-1)\oplus\mathcal{O}_{\mathbb{P}^1}(-1)$ by the $\mathbb{Z}_p$ action
(\ref{zpaction}), which corresponds to a partial resolution of $\mathcal{X}_{p,q}$. This can be described as an orbi-bundle fibration of a Hirzebruch-Jung
singularity over a $\mathbb{P}^1$ with two marked points with $\mathbb{Z}_p$-monodromy.
One way to do that is to realize the projectivization of the resolved conifold as a subspace 
$\{[z_0,z_1,z_2],[z_3,z_4,z_5],[r,s]\in \mathbb{P}^2\times\mathbb{P}^2\times\mathbb{P}^1 | z_1 r = z_2 s , z_3 r = z_4 s \}$.
The $\mathbb{Z}_p$ action on the above variables is inferred from (\ref{zpaction}) via the identification
$(z_1,z_2,z_3,z_4)=(x,-z,t,-y)$ and imposing invariance of the relations, which gives $(r,s)\to (\omega^{q-1}r,s)$.
The fiber over the north pole $r=0, s=1$ is parametrized by $(z_1,z_3)$, which describe precisely a Hirzebruch-Jung
singularity. The analogue of the above is valid for the fiber over the south pole. 
\end{remark}

All we are left to do to complete step 2 is to take a complete resolution $\overline{\mathcal{X}_{p,q}}$ of $\mathcal{X}_{p,q}$
$$\overline{\mathcal{X}_{p,q}} \stackrel{r}{\dashrightarrow} \mathcal{X}_{p,q}$$
Since $\mathcal{X}_{p,q}$  is Gorenstein, the birational morphism $r$ can be taken to preserve the condition of $\overline{\mathcal{X}_{p,q}}$ being both toric and Calabi-Yau, i.e. to be a crepant toric resolution. We can realize this diagrammatically \cite{Fulton} by adding all the interior lattice vectors inside the tetrahedron spanned by $b_i$, and declaring that the (top-dimensional part of the) fan of $\overline{\mathcal{X}_{p,q}}$ is made by the cones constructed above the 2-simplices which triangulate the projection of the fan onto the $y=1$ plane.
\begin{figure}[t]
\begin{minipage}[t]{0.49\linewidth}
\centering
\vspace{0pt}
\includegraphics[scale=0.33]{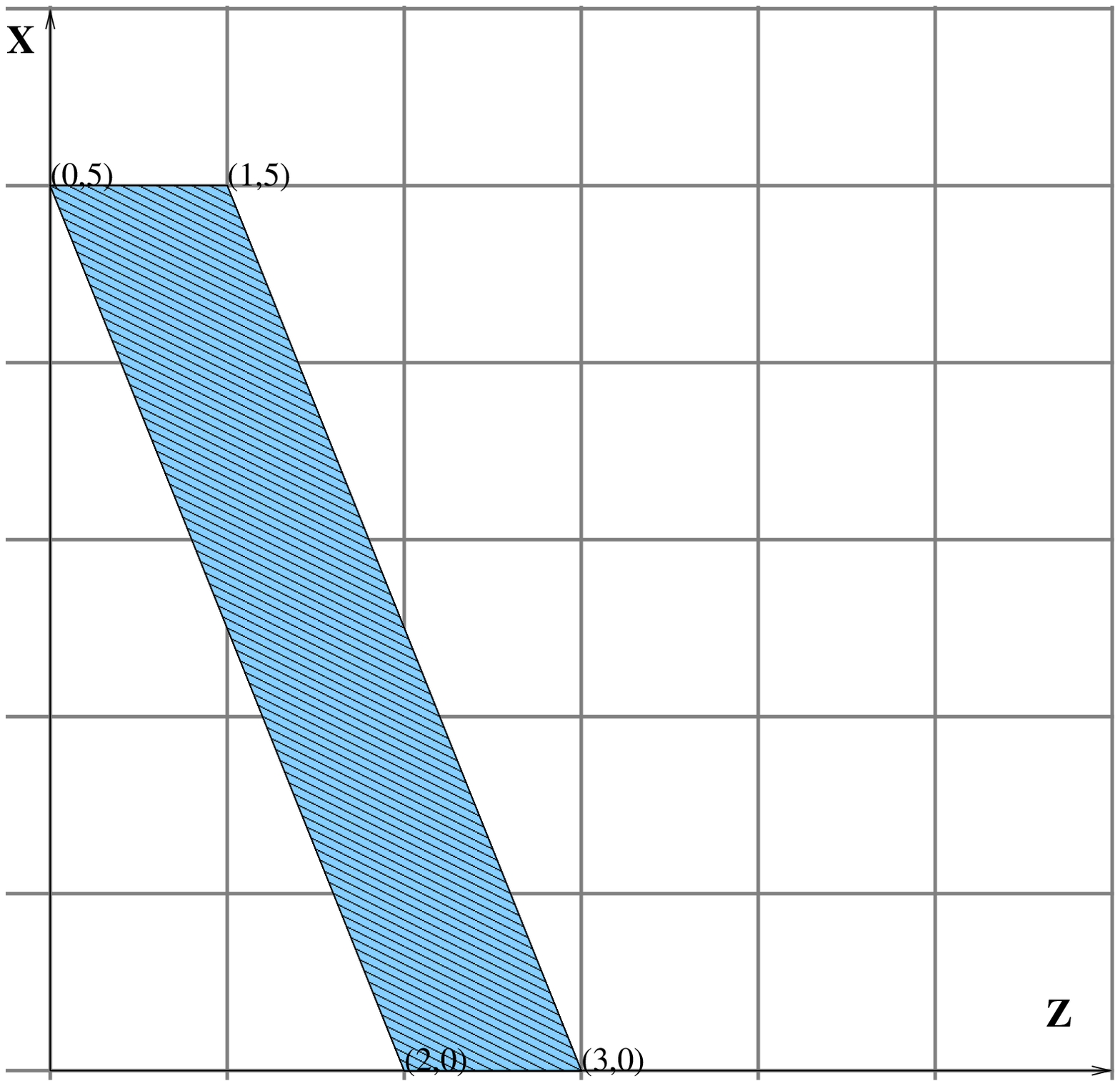}
\vspace{0pt} 
\caption{The fan of $\mathcal{X}_{p,q}$ for $p=5$, $q=2$.}
\label{conifold52}
\end{minipage}
\begin{minipage}[t]{0.49\linewidth}
\centering
\vspace{0pt}
\includegraphics[scale=0.33]{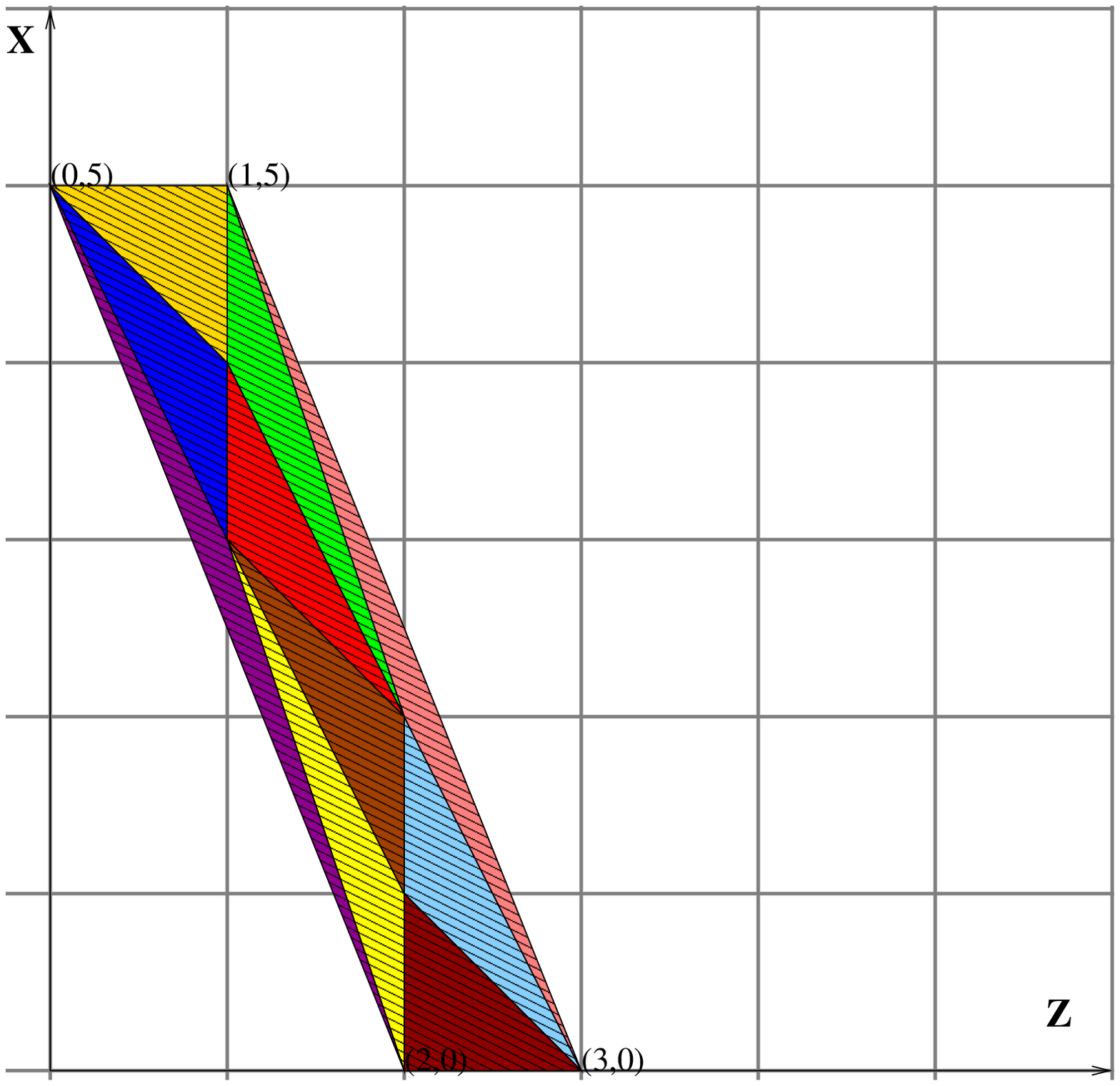}
\vspace{0pt} 
\centering
\caption{The fan of $\overline{\mathcal{X}_{p,q}}$ for $p=5$, $q=2$}\label{res52}

\end{minipage}
\end{figure}
Notice that the latter is a parallelogram with shorter sides of length 1 (see figure \ref{conifold52}). This means that the intersection of the parallelogram with the horizontal lines $x=1,\dots, p$ contains either one point of the lattice in the interior or two points on the diagonal edges, the latter possibility being excluded by the coprimality condition $(p,q)=1$. Thus the number of points of the lattice (apart from the 4 external vertices) in the interior of the parallelogram is precisely $p-1$. As is clear from the picture, these points have the form (on the plane) $([q+1 - j q/p], j)$ $=$ $(q - [j q/p], j)$, where square brackets denote the integer part of the argument.

It is straightforward to get a complete crepant resolution of the orbifold by taking a triangulation of the $p+3$ points
$$v_1\equiv \left(\begin{array}{c}q+1 \\ 0 \end{array}\right) \quad v_2\equiv \left(\begin{array}{c}q \\ 0 \end{array}\right) \quad v_{p+3}\equiv \left(\begin{array}{c}1 \\ p \end{array}\right)$$
\beq
v_{j+2}\equiv \left(\begin{array}{c} q - [j q/p] \\ j \end{array}\right), \qquad j=1, \dots, p
\label{fanres}
\eeq
\begin{defn}\label{Xpqres}
We will call $\overline{\mathcal{X}_{p,q}}$ the toric variety defined (modulo flops) by a fan supported by the rays
\beq
\label{fanpqres}
 b_i\equiv \binom{v_i}{1}
\eeq
and whose $3$-dimensional cones are defined by having their intersection with the $z=1$ hyperlane coincide with the simplices of a complete triangulation of the convex hull of (\ref{fanpqres}).
\end{defn}
By  construction $\overline{\mathcal{X}_{p,q}}$ is a simplicial, smooth\footnote{A triangulation of the $p+3$ points (\ref{fanres}) realizes the projection of our cone onto the plane $y=1$ as the disjoint union of precisely $2p$ triangles. The fact that the number of triangles is $2p$ is a consequence of Euler's formula: denoting with $m$ the number of triangles in a triangulation of (\ref{fanres}), since each triangle has $3$ edges and the convex hull has $4$, the number of edges is $(3m+4)/2$, due to the fact that each edge is incident to exactly two faces. Plugging all the ingredients (number of points and edges) into Euler's formula, it follows that the number of triangles is exactly $2p$.  Now, each triangle has half-integer area since each vertex is a site of the lattice, and the area is then given by half the determinant of an integer matrix. But $p$ is the area of the whole parallelogram, so having $2p$ triangles implies that each triangle must have area $1/2$. The cones which project onto those triangles are then simplicial and smooth (i.e. each triple of vectors spanning a cone in the fan of the resolution is an integer basis of the lattice), which is precisely the non-singularity condition for a toric variety.} toric $CY$ three-fold which is birationally isomorphic to $\mathcal{X}_{p,q}$. Step 2 is completed.

\begin{flushright}$\square$ \\ \end{flushright}

\begin{figure}[t]
\centering
\includegraphics[scale=0.20]{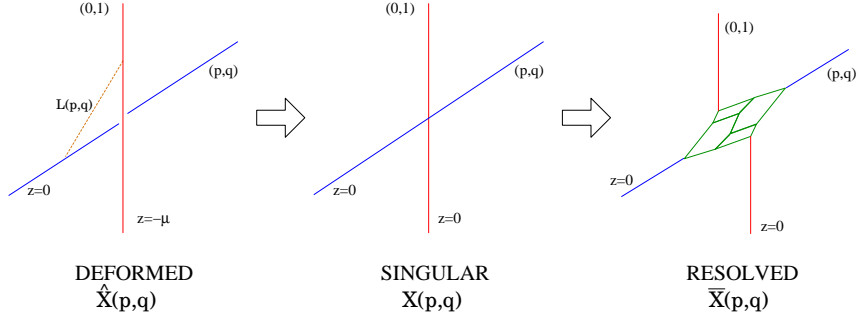}
\caption{A pictorial representation of the geometric transition for Lens spaces $L(p,q)$ as $\mathbb{T}^2$ fibrations.}
\end{figure}
\begin{figure}[t]
\centering
\includegraphics[scale=0.30]{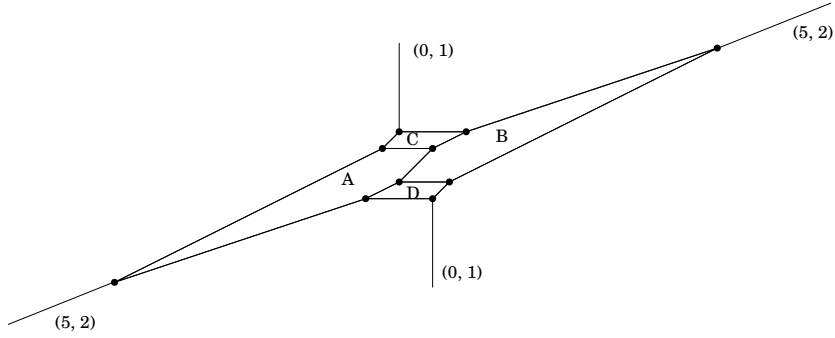}
\caption{The $pq$-web diagram for the resolution of the orbifold $p=5$, $q=2$. $A$, $B$ and $C$, $D$ represent two $dP_2$ and two $\mathbb{F}_1$ surfaces respectively.}
\label{pqweb52}
\end{figure}

The toric data (\ref{fanpqres}) allows us to extract some useful information on the geometry of $\overline{\mathcal{X}_{p,q}}$. First of all, since internal vertices are in 1-to-1 correspondence with linear equivalence classes of (compact) divisors of $\overline{\mathcal{X}_{p,q}}$, we have that the fourth Betti number is
$$b_4(\overline{\mathcal{X}_{p,q}}) = p-1$$
for every $q$. Moreover, given that the Euler characteristic $\chi(\overline{\mathcal{X}_{p,q}})$ is simply given by twice the area of the base of the tetrahedron and that odd Betti numbers vanish, we can easily compute the dimension of the second cohomology group as
\beq
b_2(\overline{\mathcal{X}_{p,q}}) = \chi(\overline{\mathcal{X}_{p,q}}) -b_0(\overline{\mathcal{X}_{p,q}}) -b_4(\overline{\mathcal{X}_{p,q}}) = 2p -1 - (p-1) = p
\eeq
for every $q$. This is expected: the dimension of the K\"ahler moduli space of $\overline{\mathcal{X}_{p,q}}$ should match the number of inequivalent flat connections of the $CS$ $SU(N)$ theory on $L(p,q)$, which is $\pi_1(L(p,q))=p$. \\
It is however remarkable, as it is also apparent from figure \ref{res52} and \ref{pqweb52}, that the intersection structure of $\overline{\mathcal{X}_{p,q}}$ for $q>1$ is significantly more complicated than the simple case $q=1$. Instead of the simple ladder diagrams describing the $pq$-webs of the $A_{p-1}$ geometries, which were built out of a single tower of (nef) Hirzebruch surfaces, the compact divisors here are generic toric Fano surfaces and intersect in a wildly more intricate way, due to the fact that the vertices of the rays of the fan are no-longer tetravalent and vertically aligned.
 \\

\subsection{Mirror symmetry}
From the toric data (\ref{fanpqres}) we can straightforwardly write down the Hori-Vafa mirror curve to $\overline{\mathcal{X}_{p,q}}$. We recall the following
\begin{defn}[Hori-Vafa mirror, \cite{Hori:2000kt}]
The $B$-model target space mirror to a toric $CY$ three-fold $X$ is the hypersurface in $\mathbb{C}^2(x_1,x_2) \times (\mathbb{C}^*)^2(U,V)$
$$x_1 x_2 = P_X(U,V)$$
where $P_X(U,V)$ is the Newton polynomial associated to the fan of $X$. Periods of the holomorphic 3-form reduce to periods of the differential
$$d\lambda_X = \log{U} \hbox{ d} \log{V}$$
on the {\bf mirror curve} $\Sigma_X$ (a projective curve with $n$ punctures) given by the zero locus $P_X(u,v)=0$. Denoting with $\mathcal{F}_X$ the projection of the fan onto the hyperplane containing the tip of the $1$-dimensional cones, the number $n$ of punctures and the genus of its $n$-point compactification are determined as
\beq
n = \hbox{\# of external points of $\mathcal{F}_X$}, \qquad g=\hbox{\# of internal points of  $\mathcal{F}_X$}
\eeq
\end{defn}
While ($\Sigma_X$, $d\lambda_X$) depends on a choice of a representation for the dual lattice of $X$, i.e. on a $GL(3,\mathbb{Z})$ equivalence class, the periods are automatically invariant. In our case we have from (\ref{fanres}), modulo a shift and a reflection on the $x$ direction, that in logarithmic coordinates $u=\log{U}$, $v=\log{V}$
\beq
\label{mirrorcurvepq}
P_{p,q}(u,v)= \left(e^{pu+qv}-1\right)\left(e^{v}-1\right)+d_p + \sum_{j=1}^{p-1}d_je^{ju+(q-[(p-j)q/p])v}
\eeq
When $d_j=0$, corresponding to the singular $\mathcal{X}_{p,q}$, this form for the mirror curve had already been suggested by \cite{Aganagic:2002wv}. Notice, from (\ref{fanres}), that there is no $GL(3,\mathbb{Z})$ transformation sending the points in $\mathcal{F}_{\overline{\mathcal{X}_{p,q}}}$ into a strip of horizontal width less than 3 for $1<q<p-1$. Moreover, by (\ref{zpaction}), the fan of $\overline{\mathcal{X}_{p,q}}$ and  $\overline{\mathcal{X}_{p,p-q}}$ are related by an automorphism of the lattice, thus yielding isomorphic toric varieties. Collecting it all together we have proven the following
\begin{prop}
\label{horivafaprop}
The Hori-Vafa mirror curve and differential are given by
$$\Sigma^{HV}_{p,q}: P_{p,q}(e^u,e^v)=0, \qquad d\lambda_{p,q}=u d v$$
where $P_{p,q}$ is given by (\ref{mirrorcurvepq}) and $u$,$v \in\mathbb{R}\times S^1$. The curve has 4 punctures and genus $p-1$ for all $q$, and its periods have a symmetry given by $q\to p-q$. For $1<q<p-1$, $\Sigma^{HV}_{p,q}$ is not hyperelliptic.
\end{prop}
\begin{flushright}$\square$\\\end{flushright}

\section{The open string side: $CS$ theory on $L(p,q)$ and matrix models}
\label{open}
The goal of this section is to provide a suitable matrix model representation of the partition function of Chern-Simons theory
on a $L(p,q)$ lens space in a given vacuum. This case has been already considered in \cite{Marino:2002fk}, where a general 
matrix integral representation for the partition function of Chern-Simons theory on Seifert homology spheres has been derived (see also 
\cite{Garoufalidis:2006ew} and \cite{Dolivet:2006ii}). Here we find a slightly different, but equivalent, representation more useful for our purposes.

More precisely, let us consider $U(N)$ Chern-Simons theory at level $k\in \mathbb{Z}$ on a $L(p,q)$ lens space ($1\leq q<p$, $p$ and $q$ coprime) and index with $\mathbf{m}\in \mathbb{Z}_p^N$ the set of $U(N)$-flat connections on $L(p,q)$. We will denote the corresponding partition function, or Reshetikin-Turaev-Witten invariant, as $Z_{U(N)}^{ L(p,q)}\big(k,\mathbf{m}\big)$. We have the following
\begin{thm}[Hansen-Takata, \cite{Hansen}]
 The Chern-Simons partition function for a $L(p,q)$ lens space, gauge group $U(N)$, level $k$ and a fixed choice of flat connection $\mathbf{m}$ is given by
\beq
Z_{U(N)}^{ L(p,q)}\big(k,\mathbf{m}\big)=\, C_N(p,q;g_s)\,
e^{- \frac{4\pi^2 q}{g_s^2p}\mathbf{m}^2}~\sum_{\tilde\omega,\omega\in S_N}\,\varepsilon(\omega)~
e^{\frac{g_s^2}{2\,p}\,\omega(\mathbf{\rho})\cdot\mathbf{\rho}}~
e^{\frac{2\pi\i}p\,\tilde\omega(\mathbf{m})\cdot(q\,\mathbf{\rho}+\omega(\mathbf{\rho}))}
\label{WUNfluctdef}\eeq
where $g_s^2=\frac{4\pi i}{k+N}=\frac{4\pi i}{\hat{k}}$, $\rho=\frac{1}{2}\sum_{\alpha>0} \alpha$ is the Weyl vector of $SU(N)$, $S_N$ is the permutation group of $N$ elements and $C_N(p,q;g_s)$ is
a fixed overall factor, not depending on the particular flat connection (the exact expression of $C_N(p,q;g_s)$ is given in \cite{Hansen} and does not play any role here).
\end{thm}
To obtain a matrix model representation it is useful to observe that this expression, up to an overall normalization factor, can be also written as
\bea
Z_{U(N)}^{ L(p,q)}\big(k,\mathbf{m}\big)&=&\sum_{\tilde\omega,\omega\in S_N}\varepsilon(\omega)~
e^{\frac{1}{4 g_s^2 p }\left(g_s^2({\omega(\mathbf{\rho} })+ \mathbf{\rho} ) +4 i \pi  \tilde\omega(\mathbf{m})\right)^2+2\pi i\frac{(q-1)}{p} \tilde
\omega(\mathbf{m})\cdot\rho}.
\eea
By exploiting a trivial integral representation of the gaussian function, we can rewrite the above partition function as an integral
\beq
\begin{split}
&Z_{U(N)}^{ L(p,q)}\big(k,\mathbf{m}\big)=\!\!\int_{-\infty}^\infty d^N x\!\!\!\!\!\sum_{\tilde\omega,\omega\in S_N}\!\!\!\!\varepsilon(\omega)~ e^{{i \hat{k} p \pi  (x\cdot x)}+
2  \pi \left({\omega(\mathbf{\rho} })+ \mathbf{\rho}  +\hat{k} ~\tilde\omega(\mathbf{m})\right)\cdot x+2\pi i\frac{(q-1)}{p} \tilde
\omega(m)\cdot\rho}\\
&=\int_{-\infty}^\infty d^N x\!\!\!\!\!\sum_{\tilde\omega,\omega\in S_N}\!\!\!\!\varepsilon(\omega)~ e^{{i \hat{k} p \pi  (x\cdot x)}+
2  \pi \left({\tilde\omega^{-1}(\omega(\mathbf{\rho} }))+ \tilde\omega^{-1}(\mathbf{\rho})  +\hat{k} ~\mathbf{m}\right)\cdot \tilde\omega^{-1}(x)+2\pi i\frac{(q-1)}{p} \mathbf{m}\cdot\tilde\omega^{-1}(\rho)}.
\end{split}
\eeq
Since the measure of integration  and $(x\cdot x)$ are symmetric under permutations, the partition function can be rearranged as
\beq
\begin{split}
&Z_{U(N)}^{ L(p,q)}\big(k,\mathbf{m}\big)=\!\!\int_{-\infty}^\infty\!\!\! d^N x\!\!\!\!\!\sum_{\tilde\omega,\omega\in S_N}\!\!\!\!\varepsilon(\omega)\varepsilon(\omega^\prime)~ e^{{i \hat{k} p \pi  (x\cdot x)}+
2  \pi \left({\omega(\mathbf{\rho} })+ \tilde\omega^{\prime}(\mathbf{\rho})  +\hat{k} ~\mathbf{m}\right)\cdot  x+2\pi i\frac{(q-1)}{p} \mathbf{m}\cdot\tilde\omega^{\prime}(\rho)}.
\end{split}
\eeq
To perform the sum over $\omega$  and $\omega^\prime$, it is sufficient to recall the Weyl-formula
\beq
\sum_{\omega\in S_N} \varepsilon(\omega) e^{i (\phi\cdot{\omega(\rho)})}=\prod_{\alpha>0}2\sin\left(\frac{\alpha\cdot \phi}{2}\right),
\eeq
and we thus get, restoring $g_s=\frac{4\pi i}{\hat k} $,
\beq
\label{MMCS}
\begin{split}
&Z_{U(N)}^{ L(p,q)}\big(k,\mathbf{m}\big)=\\
&=\int_{-\infty}^\infty d^N x~~e^{{i \hat{k} p \pi  (x\cdot x)}+2\pi \hat k~ \mathbf{m}\cdot x}\prod_{\alpha>0}\sinh\left({\pi\alpha}\cdot x
\right)
\sinh\left({\pi\alpha}\cdot \left (x+i \frac{(q-1)}{p} \mathbf{m}\right)
\right) \\
&=\int_{-\infty}^\infty d^N x~~e^{-{ g_s p  (x\cdot x)}+{4\pi} i ~ \mathbf{m}\cdot x}\prod_{i<j}
\sinh\left(\Delta_{ij}+\frac{i\pi(q-1)}{p}(\mathbf{m}_i-\mathbf{m}_j)\right)\sinh\left(\Delta_{ij}\right), \\
\end{split}
\eeq
where $\Delta_{ij}\equiv \frac{g_s}{2}(x_i-x_j)$. All the equalities hold up to irrelevant multiplicative constant factors and we have
\begin{thm} The partition function of Chern-Simons theory on a $L(p,q)$ lens space for a choice {\bf m} of flat connection can be written as a multi-eigenvalue integral as
\beq
\label{MMCS2}
\begin{array}{lcl}
Z_{U(N)}^{ L(p,q)}\big(k,\mathbf{m}\big) &=&
\int \prod_{I=1}^p d^{N_I}u^{(I)}_k e^{-\sum_{j=1}^N u_j^2 \frac{p}{2 g_s }} \prod_{i<j}\sinh\left(\hat\Delta^{(I)}_{ij}\right)\sinh\left(\hat\Delta^{(I)}_{ij}\right) \\ & & \prod_{I<J}\prod_{i<j}\sinh\left(\hat\Delta^{(IJ)}_{ij}+\frac{\pi i(I-J)}{p}\right)\sinh\left(\hat\Delta^{(IJ)}_{ij}+q\frac{\pi i(I-J)}{p}\right)
\end{array}
\eeq
where $u_i^{I}\in \mathbb{R}$, $I=1,\dots,p$, $i=1,\dots, N_I$ and we have defined $\hat\Delta^{(I)}_{ij}\equiv \frac{1}{2}\left(u_i^{(I)}-u^{(I)}_j\right)$, $\hat\Delta^{(IJ)}_{ij}\equiv \frac{1}{2}\left(u_i^{(I)}-u^{(J)}_j\right)$.
\end{thm} In (\ref{MMCS2}) we have eventually rescaled $g_s$ by a factor of two in order to make contact with the notation of \cite{Aganagic:2002wv, Halmagyi:2003mm}, to which it reduces in the case $q=1$ and discarded a constant in front of the final matrix integral. This representation 
is of course equivalent to the one found in \cite{Marino:2002fk}, up to an overall multiplicative constant.

\begin{remark} At this stage we can already spot a few signals of the fact that $GV$ duality could break down for $q>1$. Indeed, two $L(p,q)$ and $L(p',q')$ lens spaces are homeomorphic if and only if $p = p'$ and $q = \pm q' \,\,({\rm mod}\, p)$ or $qq' = \pm1 \,\,({\rm mod}\, p)$: the related partition functions are topological invariants and should thus be equal. This can be verified explicitly when the sum over the flat connections is performed and the Chern-Simons level is correctly quantized \cite{Hansen}, but the same property does not seem to show up for the partition function in the background of a fixed flat connection: as one can easily check by explicit examples, different flat connection sectors are mixed under the relevant transformations. On the other hand, as pointed out in Proposition \ref{horivafaprop}, $q = \pm q' \,\,({\rm mod}\, p)$ is instead a symmetry of the closed string background described in the the previous section. Therefore it is expected that the spectral data (\ref{mirrorcurvepq}) will be different from what we will extract from the large $N$ analysis of (\ref{MMCS2}).


\end{remark}

\subsection{Large $N$ limit of the CS matrix model}
\label{largeN}

We now would like to prove that, as in the case of hermitian matrix models, the eigenvalue integral (\ref{MMCS2}) is governed by a pair $(\Sigma^{CS}_{p,q}, d R_{p,q})$ made up of a spectral curve $\Sigma^{CS}_{p,q}$ and a resolvent $dR_{p,q}$, out of which the genus zero free energy is extracted by the usual relations of special geometry. Proving spectral equivalence as in proposition \ref{HYO} amounts then to finding an isomorphism of curves $\phi$ such that
\beq
\begin{array}{ccccc}
\phi: & \Sigma^{HV}_{p,q} & \mapsto & \Sigma^{CS}_{p,q} &\quad \hbox{isomorphism} \\
& (\phi^{-1})^* d\lambda_{p,q}& = & d R_{p,q} &
\end{array}
\label{isocurves}
\eeq
Let us first introduce some basic objects in the discussion of the large $N$ limit.
\begin{defn}
Let $N\in \mathbb{N}_0$, $I=1,\dots, p$ and $x\in \mathbb{R}$.  For every $I$, the sequence of tempered distributions $\rho^{(I)}_N \in S'(\mathbb{R}) $
$$\rho^{(I)}_N(x) := \frac{1}{N}\sum_{i=1}^N \delta(x-u^{(I)}_i)$$
will be called $\mathrm{I^{th}}$ {\rm eigenvalue density at rank $N$}. Their integral on the real line gives the relative fraction of eigenvalues {\rm (}{\bf filling fraction}{\rm )} in the $I^{th}$ group
\beq
\int_{\mathbb{R}} \rho_N(x) dx = \frac{N_I}{N}
\eeq
\end{defn}

\noindent We then make the following basic \\

\noindent {\bf Assumption.}
{\it We {\bf assume} that the $N\to\infty$ distributional limit $$\rho^{(I)}(x) := \lim_{N\to\infty} \rho^{(I)}_N(x)$$ is a compactly supported continous function on the real line, $\rho^{(I)} \in \mathcal{C}^0_c(\mathbb{R})$, $\forall I$. } \\

\noindent This assumption is motivated by the analogous situation for hermitian matrix ensembles as well as for the $CS$ matrix models of \cite{Halmagyi:2003mm}, and we will prove that it is self-consistent. It will be useful in the following to denote with $t\equiv g_s N$ the total 't Hooft coupling and with $S_I$ the large $N$ limit of the filling fractions
$$S_I := t\lim_{N\to\infty} \frac{N_I}{N}$$ normalized so that $\sum_I S_I = t$. \\

We will now construct explicitly the spectral curve $\Sigma^{CS}_{p,q}$ and differential $d R_{p,q}$ emerging from the large $N$ study of (\ref{MMCS2}). As usual in random matrix theory, this will be accomplished by finding an implicit algebraic expression $P_{p,q}(u,v)$ for the force $v(u)$ on a probe eigenvalue $u$ at large $N$, in terms of which we will define
\bea
\Sigma^{CS}_{p,q} &:=& \{ (u, v) \in (\mathbb{R} \times S^1) \times (\mathbb{R} \times S^1) | P_{p,q}(u,v)=0\} \nonumber \\
d R_{p,q} &:=& v(u) du
\eea
At large $N$, $v(u)$ will have cuts in the complex plane whose discontinuity yields the individual eigenvalue densities $\rho^I$; its regularized integral from infinity to the $I^{th}$ cut will instead measure, by construction, the variation of the (leading order) free energy with respect to the $I^{th}$ filling fraction. This is summarised by the {\it special geometry} relations
\beq
\oint_{A_I} d R_{p,q} = S_I \qquad \oint_{B_I} d R_{p,q} = \frac{\partial \mathcal{F}}{\partial S_I}
\eeq
We will now show, from the explicit form of $(\Sigma^{CS}_{p,q}, d R_{p,q})$, that no such a $\phi$ as in (\ref{isocurves}) does in fact exist for $1<q<p-1$.
\\

\noindent {\bf Proof of Claim 1.}
As is customary for $CS$ multi-matrix models, the steepest descent (saddle-point) method to evaluate (\ref{MMCS2}) at large $N$ yields a singular integral equation for $\rho^{(I)}$ with $q$-dependent hyperbolic kernels. From (\ref{MMCS2}) we can straightforwardly write the saddle point equation as
\bea
\label{saddleq}
p \lambda_I &=&
t \fpint{}{} \coth\left(\frac{\lambda_I-\lambda'_I}{2}\right)\rho_I(\lambda_I')d\lambda_I' \nonumber \\ &+& \frac{t}{2} \sum_{J \neq I} \int_\mathbb{R} \left[\coth\left(\frac{\lambda_I-\lambda_J}{2} + d_{IJ}\right)+\coth\left(\frac{\lambda_I-\lambda_J}{2}+ q d_{IJ}\right)\right]\rho_J(\lambda_J)d\lambda_J \nonumber \\
\eea
where $d_{IJ}:=i\pi (I-J)/p$ and the slashed integral indicates the Cauchy principal value (``improper'') integral. \\
Now let us define the following set of {\it resolvents}
\bea
\label{omegaIq}
\omega_I(z) &\equiv & t \int_\mathbb{R} \coth\left(\frac{z-\lambda_I}{2}\right)\rho_I(\lambda_I)d\lambda_I ,\\
\label{omegaq}
\omega(z) &=& \frac{1}{2}\sum_{I=1}^{p} \left[\omega_I \left(z-2\pi i \frac{I}{p}\right)+ \omega_I \left(z-2\pi i \frac{q I}{p}\right)\right],
\eea
\\

\noindent We will need the following easy generalization of the Sokhotski-Plemelij lemma
\begin{lem}
\label{sokplem}
Define the following limits in $S'(\mathbb{R})$
$$\coth_\pm(z) := \lim_{\epsilon \to 0} \coth\left(z\pm i\epsilon\right).$$
Then the following identities in $S'(\mathbb{R})$ hold true:
\bea
\label{sppv}
\coth_+ +\coth_-&=&2\hbox{\rm pv}(\coth), \\
\coth_+ - \coth_-&=&-2\pi i \delta.
\label{spdelta}
\eea
\end{lem}
\noindent For notational purposes, we define accordingly $\omega_{\pm}(z)\equiv\lim_{\epsilon\to 0} \omega(z+i\epsilon)$. \\

Given that the eigenvalue densities are supported on the real axis, we conclude immediately from (\ref{omegaIq}) and (\ref{spdelta}) that the individual resolvents $\omega_I(z)$ have branch cuts which coincide with\footnote{The eigenvalue integral is parity invariant, which therefore implies a $\mathbb{Z}_2$ symmetry in the location of the branch points.} $\hbox{supp} (\rho_i)=[-a_I, a_I]$ for some $a_I\in \mathbb{R}$. This implies that $\omega(z)$, as a function from the cylinder $0\leq\Im m z < 2\pi$ to the Riemann sphere, has $p$ cuts centered at $z=2\pi i I/p$, whose width as usual depends on a choice of filling fractions $S_I$.  Explicitly, from (\ref{omegaq}) we have for $J=0,\dots, p-1$
\beq
2\omega\left(z+\frac{2\pi i J}{p}\right)=\omega_J\left(z\right)+\omega_{\hat J}(z)+\sum_{I\neq J}\omega_I\left(z-\frac{2\pi i (I-J)}{p}\right) + \sum_{I\neq \hat J} \omega_{I}\left(z-\frac{2\pi i (qI-J)}{p}\right)
\eeq
where  $\hat J$ is defined by $  q \hat J =J \hbox{ mod } p$. When $I=0$ we have $\hat I=0$, and for $x\in[-a_0, a_0]$ we get that
\beq
\begin{array}{lcl}
\frac{\omega_+(x)+\omega_-(x)}{2}&=&
t \fpint{}{} \coth\left(\frac{x-\lambda'_I}{2}\right)\rho_0(x')dx'  \\ &+& \frac{t}{2} \sum_{J \neq 0} \int_\mathbb{R} \left[\coth\left(\frac{x-x'}{2} + d_{pJ}\right)+\coth\left(\frac{x-x'}{2}+ q d_{pJ}\right)\right]\rho_J(x')dx'  \\
&=& p x
\end{array}
\eeq
due to (\ref{saddleq}) and lemma \ref{sokplem}. However, a quick inspection shows that for no other $0<I<p$ it is possible to find a closed expression for the average of the resolvent on the $I^{th}$ cut. Indeed, since $I\neq \hat I$ for $0<I<p$, different individual resolvents become singular at $x+2\pi i I/p$ inside the total sum (\ref{omegaq}), namely $\omega_I$ and $\omega_{\hat I} \neq \omega_I$, and it appears to be very intricate to infer the structure of $\omega(z)$ from (\ref{saddleq}). However, let us  restrict ourselves for the moment to the special case in which
\beq
\rho_I = \rho_1 \qquad  1<I<p
\label{Iuguali}
\eeq
This corresponds to a particular symmetric choice of filling fractions, that is one in which a fraction of $S_0$ eigenvalues have been put on the cut on the real axis, corresponding to the trivial Chern-Simons connection, and $S_I=(t-S_0)/(p-1)$ for $I>0$ are democratically distributed between the non-trivial flat connections. This would amount to explore a peculiar codimension $p-2$ subspace in the space of 't Hooft parameters, for which the large $N$ data can be described in complete detail. In particular, this would give a 2-parameter closed subset of our sought-for $p$-dimensional family $(\Sigma_{p,q}^{CS}, d R_{p,q})$. \\
Under the constraints (\ref{Iuguali}) we now have that, for $x\in [-a_I, a_I]$,
\beq
\frac{\omega_+\left(x+\frac{2\pi i I}{p}\right)+\omega_-\left(x+\frac{2\pi i I}{p}\right)}{2}= p x  \qquad \forall  I=0,\dots, p-1
\label{saddleenh}
\eeq
Now, let's map conformally the cylinder of width $2\pi$ to the punctured complex plane via
\beq
\begin{array}{cccc}
Z: &\mathbb{R} \times  S^1 & \mapsto & \mathbb{C}^* \\
& z &\mapsto & e^{z}.
\end{array}
\eeq
Exponentiating (\ref{saddleenh}) yields
\beq
Z^p = e^{\omega_+/2}e^{\omega_-/2}
\eeq
Introduce now
\beq
g(z)=e^{\omega/2}+Z^pe^{-\omega/2}
\label{g(z)}
\eeq

\noindent This is a function which is single-valued on the whole strip  $0\leq\Im m z < 2\pi$, because for all $x\in [-a_I, a_I]$ we have that
\bea
g_+(x+2\pi i I/p )&=& e^{\omega_+/2}+e^{p(x+2\pi i I/p)}e^{-\omega_+/2} =  e^{\omega_+/2} + e^{px}\nonumber \\ &\times & e^{-\omega_+/2}  = e^{px}e^{-\omega_-/2} + e^{\omega_-/2} = g_-(x+2\pi i I/p). 
\eea
and it is regular everywhere, except perhaps at infinity.
This implies that $g(Z)$ is an entire analytic function in $Z$ with algebraic growth
\beq
g(Z)=\sum_{n=0}^{p}d_nZ^n
\eeq
where the $d_n$'s are (still unknown) functions of the two filling fractions $S_0$, $S_I=S_1=(t-S_0)/(p-1)$.
The resolvent $\omega(z)$ is then determined by solving the quadratic equation (\ref{g(z)}) with the appropriate boundary condition at infinity, which yields
\beq
\omega(z)=\log\left[\frac{1}{2}\left(g(z)-\sqrt{g^2(z)-4 e^{pz}}\right)\right].
\eeq
Defining $u\equiv z$, $v\equiv(t-2\omega)$ we arrive at the following form for $\Sigma_{p,q}^{CS}$ under the constraint\footnote{We also set $d_1=d_p=1$ with a redefinition $u$ and dividing by an overall factor.} (\ref{Iuguali})
$$
\Sigma_{p,q}^{CS}: \qquad e^{t-2v} - e^{t/2-v}e^{-t/2}\left(e^{pu}+\sum_{n=1}^{p-1}d_ne^{nu} + 1\right) + e^{pu}=0 \Rightarrow
$$
\beq
\label{mirrorq=1}
e^{t} - e^{v}\left(e^{pu}+\sum_{n=1}^{p-1}d_ne^{nu} + 1\right) + e^{pu+2v}= (e^{v}-1)(e^{pu+v}-1) + e^t - 1 + e^v\sum_{n=1}^{p-1}d_ne^{nu}=0,
\eeq
which coincides with the Hori-Vafa mirror curve (\ref{mirrorcurvepq}) for $q=1$ and a proper identification of the complex structure parameters. We have then proven the following
\begin{prop} Let $\Sigma_{p,q}^{CS}$ and $d R_{p,q}\equiv v du$ be the 2-parameter family (\ref{mirrorq=1}) of large $N$ spectral curves and differentials of the $L(p,q)$ Chern-Simons matrix model (\ref{MMCS2}) under the constraint (\ref{Iuguali}). Then they are $q$-independent and they make up a closed subset of the family of Hori-Vafa mirror curves (\ref{mirrorcurvepq}) with $q=1$.
\end{prop}

This concludes the proof of Claim 1 for the following reason. Notice that this restricted class of $L(p,q)$ large $N$ curves consists of hyperlliptic Riemann surfaces, since they coincide with the large $N$ curves of the $q=1$ case. Moreover, they are generically smooth and have topological genus $p-1$. But for $1<q<p-1$, there is {\it no} such a subfamily inside the Hori-Vafa family of mirror curves (\ref{mirrorcurvepq}), as follows from the discussion preceding Proposition \ref{horivafaprop}.\footnote{Indeed, the very definition of the Hori-Vafa map and the fact that the toric diagram is not contained in a strip of width at most two for $1<q<p-1$  imply that hyperellipticity can be obtained only imposing a vanishing condition on a coefficient multiplying a monomial associated to an external point in the toric diagram. This amounts to discard a 1-dimensional ray in the fan and all the three-dimensional cones in which it is contained, as is familiar from the degenerate limit in which local surfaces reduce to local curves. But this fact will automatically lower the number of internal points and thus the genus of the mirror curve. Hence there can be no hyperelliptic {\it and} genus $p-1$ subfamily of curves inside (\ref{mirrorcurvepq}).}  \begin{flushright}$\square$\end{flushright}

\begin{figure}[!h]
\begin{minipage}[t]{0.49\linewidth}
\centering
\includegraphics[scale=0.25]{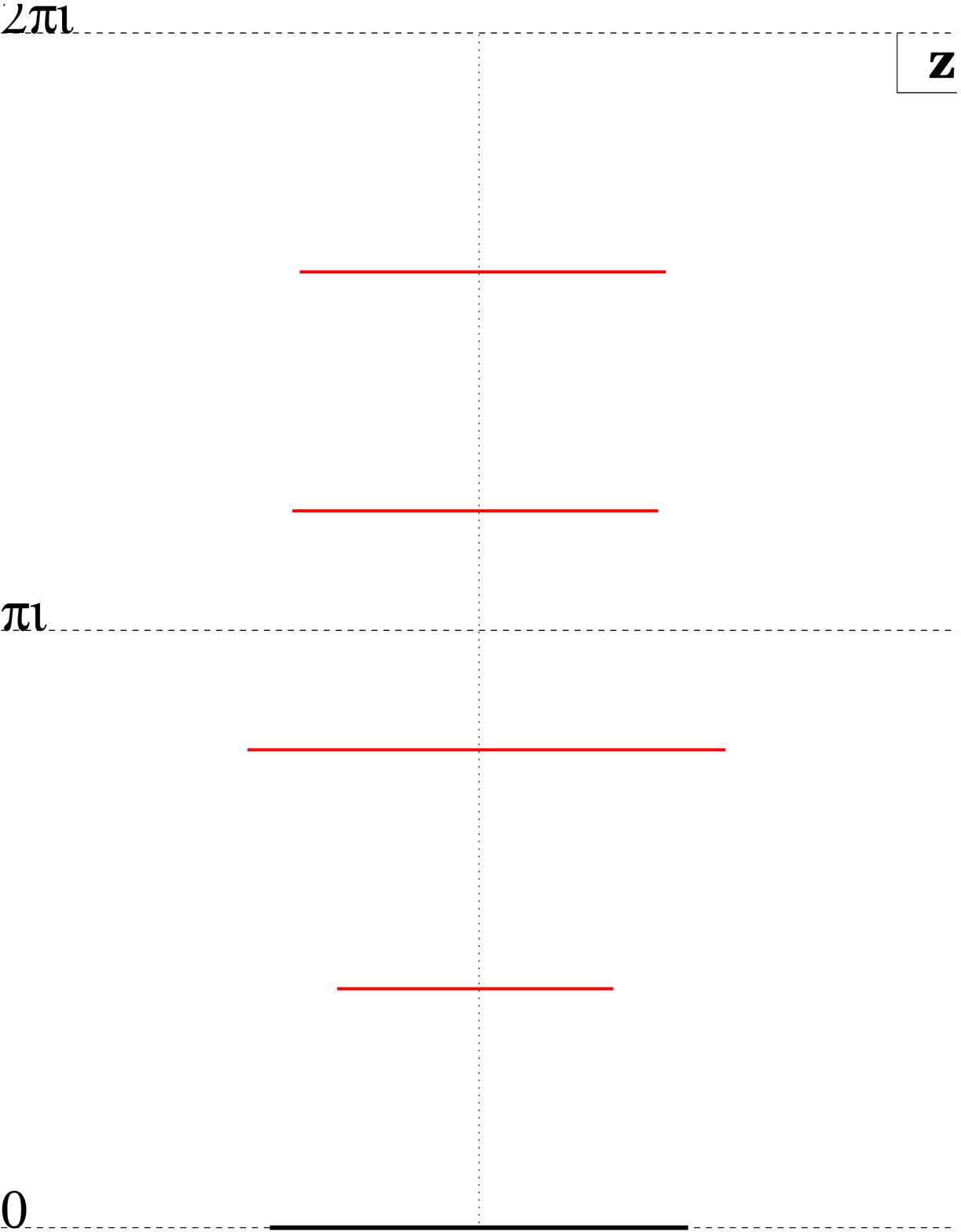}
\small
\caption{\small Cuts of the resolvent for $p=5$. Cuts relative to non-trivial flat connections  are drawn in red.}
\label{cuts52}
\end{minipage}
\begin{minipage}[t]{0.49\linewidth}
\centering
\includegraphics[scale=0.25]{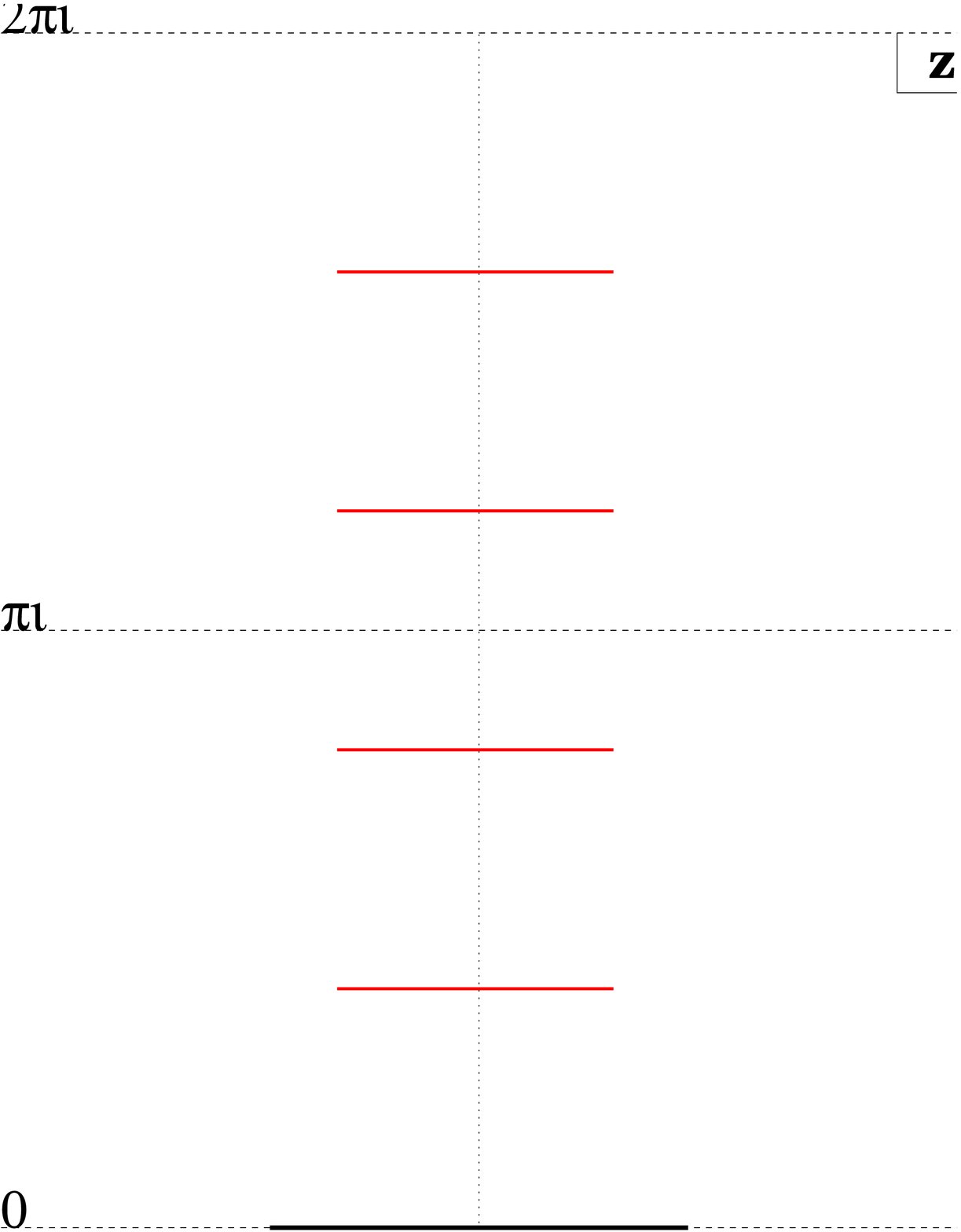}
\centering
\small
\caption{\small Cuts of the resolvent for $p=5$, imposing the constraint (\ref{Iuguali}).  Cuts relative to non-trivial flat connections  are drawn in red.}
\label{cuts52c}
\end{minipage}
\end{figure}

\vspace{1cm}
{\bf Acknowledgements. } We are particularly grateful to Etienne Mann for his advice on the content of section 2. A. B. would like to thank Boris Dubrovin, Jarah Evslin, Barbara Fantechi, Davide Forcella, Tamara Grava, Luca Philippe Mertens, Paolo Rossi and Sara Pasquetti for useful discussions. We are moreover indebted to Marcos Mari\~no for his suggestions and comments on the content of section 3. The present work is partially
supported by the European Science Foundation Programme ``Methods
of Integrable Systems, Geometry, Applied Mathematics'' (MISGAM) and
Marie Curie RTN ``European Network in Geometry, Mathematical Physics
and Applications'' (ENIGMA).

\appendix

\section{CSMM as a unitary matrix model}
An alternative matrix model realization can be given, in the trivial vacuum $m=0$, as an integral over unitary matrices.
Starting from (\ref{WUNfluctdef}) and reasoning along the lines of \cite{Marino:2002fk} we have that the $CS$ partition function can be written as
\beq
\label{MMCS1a}
Z_{U(N)}^{ L(p,q)}\big(k,\mathbf{0}\big) =\int d^N x e^{-\frac{p(x \cdot x)}{g_s q}}\prod_{i<j}\sinh\left(\frac{x_i-x_j}{2q}\right)\sinh\left(\frac{x_i-x_j}{2}\right),
\eeq
and by means of the identity
\bea
\prod_{i<j}\sinh\left(a(x_i-x_j)\right)
&=& \frac{e^{-a(N-1)\sum_{i}x_i}}{2^{N(N-1)/2}}\Delta(e^{2ax_i}),
\eea
we can write (\ref{MMCS1a}) as follows
\beq
Z_{CS}=\int_{-\infty}^{+\infty}\frac{d^N x}{2^{N(N-1)/2}} \exp\left[\frac{-p(x\cdot x)}{g_s q}-\frac{(N-1)(q+1)}{2q} \sum_i x^i\right]\Delta(x_i)^2 \frac{\Delta (e^{x_i/q})\Delta(e^{x_i})}{\Delta(x_i)^2}.
\eeq
Now, with the help of  the Itzykson-Zuber formula \cite{Itzykson:1979fi}
\bea
\frac{\det(e^{jx_i/q})}{\Delta(i^j) \Delta(x_i^j)}&=&\frac{1}{\prod_{p=0}^{N-1}p!}\left(\frac{1}{q}\right)^{\frac{N(N-1)}{2}}\int dU_1 e^{\frac{1}{q}\tr(U_1A_DU_1^\dagger X_d)}
\eea
with $A_D=\diag\left(1,\dots,N\right)$ and using
that
\beq
\int_{\mathfrak{u}(N)}f(X)d X = \Omega_N \int_{\mathbb{R}^N}d^N x \Delta^2(x) f(\hbox{diag}(x_i))
\eeq
for any Ad-invariant $f:\mathfrak{u}(N)\to \mathbb{C}$, where $\Omega_N=(2\pi)^{N(N-1)/2}/\prod_{j=1}^{N}j!$, we can turn (\ref{MMCS}) for $\mathbf{m}=0$ into a HUU 3-matrix integral
\bea
Z_{CS}&=&\left(\frac{N!}{(4\pi q)^{N(N-1)/2}}\right)\int d X e^{\frac{-p}{g_s q}\tr X^2-\frac{(N-1)(q+1)}{2q} \tr X} \nonumber \\
&\times & \int dU_1 dU_2 e^{\frac{1}{q}\tr(U_1A_DU_1^\dagger X) + \tr(U_2 A_D U_2^\dagger X)}
\eea
Defining $\hat X \equiv U_1^\dagger X U_1$, $U\equiv U_1^\dagger U_2$ and exploiting the translation invariance of the Haar measure on $U(N)$
\bea
Z_{CS}&=&\left(\frac{N!}{(4\pi q)^{N(N-1)/2}}\right) \int d\hat X e^{\frac{-p}{g_s q}\tr\hat X^2-\frac{(N-1)(q+1)}{2q} \tr\hat X} \nonumber \\
&\times & \int dU e^{\frac{1}{q}\tr(A_D \hat X) + \tr(U A_D U^\dagger X\hat)}
\eea
The gaussian integral over $\hat X$ gives
\bea
Z_{CS}&=& \left(\frac{N!}{(4\pi q)^{N(N-1)/2}}\right)\int dU e^{\frac{g_s}{2p}\tr UAU^\dagger A} \nonumber \\ & & \exp\left\{\frac{g_s}{8pq}\left[\left(\frac{N-1}{4}\right)^2N(q+1)^2+(q^2+1)\hbox{Tr}A^2 - (N-1)(q+1)^2\hbox{Tr} A\right]\right\}
\eea
Notice that this  last integral can be explicitly evaluated  by means of the Itzykson-Zuber formula \cite{Itzykson:1979fi}.
The result is in agreement with \cite{Dolivet:2006ii}, where the same expression was computed by exploiting   the  biorthogonal
polynomials.


\vskip 1truecm
\begin{thebibliography}{99}

\bibitem{Aganagic:2002wv}
  M.~Aganagic, A.~Klemm, M.~Marino and C.~Vafa,
  ``Matrix model as a mirror of Chern-Simons theory,''
  JHEP {\bf 0402}, 010 (2004)
  [arXiv:hep-th/0211098].

\bibitem{Auckly:2007zw}
  D.~Auckly and S.~Koshkin,
  ``Introduction to the Gopakumar-Vafa large $N$ duality,''
  Geometry \& Topology Monographs {\bf 8} (2006) 195-456
  [arXiv:math/0701568].


\bibitem{Bouchard:2007ys}
  V.~Bouchard, A.~Klemm, M.~Marino and S.~Pasquetti,
  ``Remodeling the B-model,''
  arXiv:0709.1453 [hep-th].

\bibitem{Cox:2000vi}
  D.~A.~Cox and S.~Katz,
  ``Mirror symmetry and algebraic geometry,''
{\it  Providence, USA: AMS (2000) 469 p}

  \bibitem{Dolivet:2006ii}
  Y.~Dolivet and M.~Tierz,
  ``Chern-Simons matrix models and Stieltjes-Wigert polynomials,''
  J.\ Math.\ Phys.\  {\bf 48}, 023507 (2007)
  [arXiv:hep-th/0609167].



\bibitem{Eynard:2007hf}
  B.~Eynard, M.~Marino and N.~Orantin,
  ``Holomorphic anomaly and matrix models,''
  JHEP {\bf 0706}, 058 (2007)
  [arXiv:hep-th/0702110].


\bibitem{Eynard:2008yb}
  B.~Eynard,
  ``Large N expansion of convergent matrix integrals, holomorphic anomalies,
  and background independence,''
  arXiv:0802.1788 [math-ph].


\bibitem{Eynard:2008he}
  B.~Eynard and M.~Marino,
  ``A holomorphic and background independent partition function for matrix
  models and topological strings,''
  arXiv:0810.4273 [hep-th].

\bibitem{Fulton}
W.~Fulton, ``Introduction to toric varieties,'' Annals of Mathematics Studies, {\bf 131}. The William H. Roever Lectures in Geometry. Princeton University Press
\bibitem{Garoufalidis:2006ew}
  S.~Garoufalidis and M.~Marino,
  ``On Chern-Simons Matrix Models,''
  arXiv:math/0601390.

\bibitem{mirror}
A.~B.~Givental, ``A mirror theorem for toric complete intersections'', in Topological field theory, primitive forms and related
topics (Kyoto, 1996), pp. 141-175; Progr. Math., {\bf 160}. Birkh\"auser Boston, Boston, MA, 1998;
B.~H.~Lian, K.~Liu, and S.~T.~Yau, ``Mirror principle I'', Asian J. Math., {\bf 1} (1997), 729-763;
A.~Elezi, ``Mirror symmetry for concavex vector bundles on projective spaces'', Int\ J\ Math.\ Math.\ Sci.\ (2003),
159-197.


\bibitem{Gopakumar:1998ki}
  R.~Gopakumar and C.~Vafa,
  ``On the gauge theory/geometry correspondence,''
  Adv.\ Theor.\ Math.\ Phys.\  {\bf 3} (1999) 1415
  [arXiv:hep-th/9811131].

\bibitem{Grassi:2002tz}
  A.~Grassi and M.~Rossi,
  ``Large N dualities and transitions in geometry,''
  arXiv:math/0209044,   


\bibitem{Halmagyi:2003mm}
  N.~Halmagyi, T.~Okuda and V.~Yasnov,
  ``Large N duality, lens spaces and the Chern-Simons matrix model,''
  JHEP {\bf 0404} (2004) 014
  [arXiv:hep-th/0312145].
  N.~Halmagyi and V.~Yasnov,
  ``The spectral curve of the lens space matrix model,''
  arXiv:hep-th/0311117.

\bibitem{Hansen}
S.~K.~Hansen, T.~Takata,
``Reshetikhin-Turaev invariants of Seifert 3-manifolds for classical simple Lie algebras'', J. Knot Theory Ramifications {\bf 13} (2004)
[arXiv:math/0209403],

  L.~Griguolo, D.~Seminara, R.~J.~Szabo and A.~Tanzini,
  Nucl.\ Phys.\  B {\bf 772} (2007) 1
  [arXiv:hep-th/0610155].

\bibitem{Hori:2000kt}
  K.~Hori and C.~Vafa,
  ``Mirror symmetry,''
  arXiv:hep-th/0002222,

  B.~Feng, Y.~H.~He, K.~D.~Kennaway and C.~Vafa,
  ``Dimer models from mirror symmetry and quivering amoebae,''
  arXiv:hep-th/0511287.

\bibitem{Itzykson:1979fi}
  C.~Itzykson and J.~B.~Zuber,
  ``The Planar Approximation. 2,''
  J.\ Math.\ Phys.\  {\bf 21} (1980) 411.

\bibitem{Marino:2002fk}
  M.~Marino,
  ``Chern-Simons theory, matrix integrals, and perturbative three-manifold
  invariants,''
  Commun.\ Math.\ Phys.\  {\bf 253}, 25 (2004)
  [arXiv:hep-th/0207096].

\bibitem{Marino:2007te}
  M.~Marino, R.~Schiappa and M.~Weiss,
  ``Nonperturbative Effects and the Large-Order Behavior of Matrix Models and
  Topological Strings,''
  arXiv:0711.1954 [hep-th].

\bibitem{Marino:2008ya}
  M.~Marino,
  ``Nonperturbative effects and nonperturbative definitions in matrix models
  and topological strings,''
  arXiv:0805.3033 [hep-th].

\bibitem{Ooguri:2002gx}
  H.~Ooguri and C.~Vafa,
  ``Worldsheet derivation of a large N duality,''
  Nucl.\ Phys.\  B {\bf 641}, 3 (2002)
  [arXiv:hep-th/0205297];
  T.~Okuda and H.~Ooguri,
  ``D branes and phases on string worldsheet,''
  Nucl.\ Phys.\  B {\bf 699}, 135 (2004)
  [arXiv:hep-th/0404101].

\bibitem{Rossi:2004eq}
  M.~Rossi,
  ``Geometric transitions,''
  J.\ Geom.\ Phys.\  {\bf 56} (2006) 1940
  [arXiv:math/0412514].
\bibitem{Stiefel}
E.~Stiefel, ``Richtungsfelder und Fernparallelismus in $n$-dimensionalen Mannigfaltigkeiten'',  Comment. Math. Helv.  {\bf 8}  (1935),  no. 1, 305--353.

\bibitem{'tHooft:1973jz}
  G.~'t Hooft,
  ``A planar diagram theory for strong interactions,''
  Nucl.\ Phys.\  B {\bf 72} (1974) 461.



\bibitem{Witten:1988hf}
  E.~Witten,
  ``Quantum field theory and the Jones polynomial,''
  Commun.\ Math.\ Phys.\  {\bf 121} (1989) 351.
\bibitem{Witten:1992fb}
  E.~Witten,
  ``Chern-Simons Gauge Theory As A String Theory,''
  Prog.\ Math.\  {\bf 133} (1995) 637
  [arXiv:hep-th/9207094].


\end{thebibliography}
\end{document}